# Mobile Cloud Forensics: An Analysis of Seven Popular Android Apps[1]

Ben Martini; Quang Do; Kim-Kwang Raymond Choo

Information Assurance Research Group, University of South Australia

ben.martini@unisa.edu.au; quang.do@mymail.unisa.edu.au; raymond.choo@unisa.edu.au

**Abstract**

Using the evidence collection and analysis methodology for Android devices proposed by Martini, Do and Choo (2015), we examined and analyzed seven popular Android cloud-based apps. Firstly, we analyzed each app in order to see what information could be obtained from their private app storage and SD card directories. We collated the information and used it to aid our investigation of each app's database files and AccountManager data. To complete our understanding of the forensic artefacts stored by apps we analyzed, we performed further analysis on the apps to determine if the user's authentication credentials could be collected for each app based on the information gained in the initial analysis stages. The contributions of this research include a detailed description of artefacts, which are of general forensic interest, for each app analyzed.

## 1  INTRODUCTION

This chapter provides a number of proof-of-concept implementations of the collection and analysis methodology described in the preceding chapter (see [1]).

We implemented the methodology using a selection of popular free cloud apps. We selected apps in the categories of storage, note taking and password sync. For the storage category, we selected Dropbox (version 2.4.1), OneDrive (version 2.5.1), Box (version 3.0.2), and ownCloud (version 1.5.5). We also investigated UPM (Universal Password Manger version 1.15) to understand the collection requirements for a third-party app which uses a cloud storage product. For the note taking category, we selected Evernote (version 5.8.1) and OneNote (version 15.0.2727.2300).

The research was conducted between April and July 2014. At the time of this research, findings are accurate to the best of the authors' knowledge. However, new releases of mobile apps and operating systems may change the way the data is collected and analyzed in the future.

# 2 ANDROID CLOUD APPS

## 2.1 DROPBOX

Dropbox is one of the top downloaded Android apps with over 100 million downloads on the Google Play store at the time of research. Dropbox provides cloud-based storage services compatible with a range of devices including PCs and mobile devices.

The following subsections describe our findings for the examination and analysis stages of the collection and analysis methodology described in [1]. Further analysis of this app is discussed in Section 2.8.1. For Dropbox:

- The **private app storage path** is "/data/data/com.dropbox.android"
- The **external storage data path** is "/sdcard/Android/data/com.dropbox.android"

### 2.1.1 Examine App Files in Private Storage (Dropbox)

Using the methodology described in [1], we located and analyzed the miscellaneous files contained within the app's private directory on the Android device's internal storage. It should be noted that this section does not aim to provide an exhaustive listing of all of the files/directories located in the apps private storage, rather it only discusses items which we believe would be of particular interest to forensic practitioners.

Dropbox stores a number of items of interest in its "dropbox-credentials.xml" shared preferences file stored at "[**private app storage path**]/shared_prefs/dropbox-credentials.xml". The file is encoded in a standard plain text XML format. The following directives of interest were located within this file:

- **app_key** – The app_key is a necessary component of the token used to authenticate the user to the Dropbox servers. It is represented using an XML String tag. The app key is the OAuth 1.0 consumer key which represents the app (rather than the user) to the Dropbox authentication system. This is discussed further in Section 2.8.1.
- **accounts** – The accounts string contains a JSON formatted value which encloses the following values of interest.
    - **userToken** – The userToken value contains two strings delimited by a pipe, both of which form part of the authentication headers used by Dropbox.
        1. The first string is the "oauth_token" which as the name suggests is the user's token. The token is updated by the app when the user initially authenticates (using a username and password) and when the previous token expires.
        2. The second string forms the second half of the "oauth_signature", which is also used as part of the authentication header in the format "[String]&[String]". This is discussed further in Section 2.8.1.
    - **userId** – This string stores the user's numerical ID, which is used in databases and the Android filesystem.

### 2.1.2 Examine App Files on External Storage (Dropbox)

After examining the miscellaneous files contained within the app's private directory, we examined the collected files from the external storage. In the case of Dropbox, we found that the following paths were populated.

**[external storage data path]/cache/[User ID]/miscthumbs**
- This directory is populated with a 'journal' file containing (mostly numerical) data, which appears to relate to the layout of the thumbnail cache.

**[external storage data path]/cache/[UserID]/thumbs/[subdirectories if any]/**

- This directory contains subdirectories named for the path of the image files stored in the user's Dropbox. e.g. "thumbs/My Pictures/san francisco.jpg".

**[external storage data path]/cache/[User ID]/thumbs/[subdirectories]/[image name].[extension]**
- The directory, named after the user's image, contains a number of sizing variations of the image presumably for display in different parts of the app.
- An interesting anomaly we noted during our research was that when images are moved on the mobile device between directories in the user's Dropbox app, the previous thumbs path and images remain. The thumbnails at the original path were not deleted even when the image was deleted from Dropbox; however, the thumbnails were deleted for the new path.

**[external storage data path]/cache/[User ID]/tmp**
- During our experiments, we noted that this directory is used to store temporary files (of the filename format "file0.tmp") as they are being downloaded. Once the download is completed, the file is moved to the scratch directory and the tmp file is deleted.

**[external storage data path]/files/[User ID]/scratch**
- This directory contains the local cache of files that have been downloaded by a user via Dropbox. These files are stored unmodified in this directory with their original filenames intact.

### 2.1.3 Examine App Databases (Dropbox)

While there are a number of possible database formats, apps generally store data in the SQLite format (presumably due to Android's native support for the format). Dropbox stores a number of SQLite databases in its private storage "[**private app storage path**]/databases/" directory. This directory contained two database files of particular interest as well as two others of lesser interest. The latter were ("[User ID]-prefs.db" and "prefs.db") and the values of interest within them (such as "Account") were encoded or encrypted. The two databases of interest were "prefs-shared.db" and "[User ID]-db.db" and are described further in Table 1 and Table 2. In addition to these database files, Dropbox stores another cache database in the "[**private app storage path**]/app_DropboxSyncCache/[app_key]" location. This database file is described in Table 3.

| [private app storage path]/databases/prefs-shared.db | |
|---|---|
| "DropboxAccountPrefs" Table | |
| **Attribute** | **Description** |
| LAST_REPORT_HOST_TIME | A millisecond resolution timestamp (in UTC) for the last report between the Dropbox app and server. This attribute appears to use the device clock rather than remote time. |

*Table 1 "prefs-shared.db" Database File*

| [private app storage path]/databases/[user-id].db | |
|---|---|
| "dropbox" Table (describes the files stored in Dropbox) | |
| **Attribute** | **Description** |
| modified_millis | The timestamp in millisecond resolution of the last modification to the file on the Dropbox server. |
| bytes | The size of the file in bytes. |
| revision | An encoding presumably of the file revision for sync purposes. The first digit increments when the file is modified on the client or server. The last 8 characters of the string match other files in the same directory. |
| hash | Only folders had a 'hash' in our experiments. While it appears to be an MD5 hash, it did not match the folder path string or other similar permutations. |
| is_dir | This field is 1 if the entry is a directory and 0 if is not (i.e. Boolean representation). |
| path & canon_path | The path (including the name) to the file/directory from the root directory; path maintains capitalization whereas canon_path is lowercase. |

| Field | Description |
|---|---|
| mime_type | The MIME type of the file. |
| thumb_exists | This field indicates if a thumbnail has been cached in a Boolean representation. |
| Parent_path & canon_parent_path | The parent path for the item with similar representation to path & canon_path. |
| _display_name | The file/directory name and extension. |
| Is_favorite | If the item is marked as a 'favorite' in Dropbox, a 1 will be entered in this field (Boolean representation). |
| local_modified | A timestamp with millisecond resolution will be entered in this field representing the modified time for the file on the local file system (field is blank if not cached). |
| local_revision | The revision code (as discussed above) for the cached copy of the file. |
| local_hash | An MD5 hash of the cached copy of the file. |
| "albums" Table (describes the image albums created by the user) | |
| col_id | The album's unique identifier (22 characters). |
| name | The album's friendly name (as entered by the user). |
| count | The number of items in the album. |
| cover_image_canon_path | The path to file for the image used as album cover. |
| share_link | The full URL for public sharing link for the Album (if user enabled). |
| creation_time | A timestamp with millisecond resolution representing the time the album was created. |
| update_time | A timestamp with millisecond resolution representing the time the album was last updated. |
| "album_item" Table (describes the album items) | |
| col_id | The album identifier to which the item belongs as recorded in "albums" table. |
| item_id | The item's unique identifier (22 characters). |
| canon_path | The path from the root directory to the file (including the item filename and extension). |
| "camera_upload" Table (describes the camera upload operations) | |
| local_hash | This "hash" is a concatenation of the file size in bytes, a "/" and the filename including extension. |
| server_hash | Presumably an MD5 hash of the file which is uploaded to the server. We checked this hash against the file in our DCIM directory and found that it did not match. |
| uploaded | A Boolean representation of whether the file has been uploaded. |
| "pending_upload" Table (describes the pending upload operations) | |
| class | This field describes the type of upload which is pending (e.g. UploadTask for files or CameraUploadTask for DCIM images). |
| data | JSON encoded key-value pairs describing the items waiting to be uploaded. The values vary depending on the class. For example, CameraUploadTask keys include mImportTimeoffset, mMimeType, mFilePath and mImportTime; and UploadTask keys include mLocalUri, mDropboxDir and mDestinationFilename. |
| "photos" Table (describes the photos in the user's Dropbox) | |
| item_id | The item's unique identifier (22 characters). |
| canon_path | The path from the root directory to the file (including the item filename and extension). |
| time_taken | A timestamp with millisecond resolution representing the time the image was taken. |
| "thumbnail_info" Table (describes the thumbnails cached by Dropbox) | |
| dropbox_canon_path | The path from the root directory to the file (including the item filename and extension). |
| thumb_size | A human readable description of the thumbnail size (e.g. "1024x768_bestfit"). |
| revision | Presumably a record of the revision of the original file from which the thumbnail was taken. |

*Table 2 "[user-id].db" Database File*

| [private app storage path]/app_DropboxSyncCache/[app_key]/[User ID]-notifications |  |
|---|---|
| "kv" Table | |
| **Attribute** | **Description** |
| app_key | The static app key used by Dropbox as the "oauth_consumer_key" in the header of the URL request when authenticating the user. It represents the app that is communicating with the Dropbox servers rather than the user. |

*Table 3 "[User ID]-notifications" Databases File*

### 2.1.4 Examine and Analyze Accounts Data (Dropbox)

As Dropbox stores the majority of its authentication information within the Dropbox app's private directory, it does not store significant data within the device's AccountManager API. After replacing the services.jar file as per the Android data collection process described in [1], we were able to obtain (via our appropriately signed app) the user's email address from within the AccountManager API. The password field for the Dropbox account, in our case, was empty.

## 2.2 BOX

Box is a popular file-syncing storage service with over 5 million downloads on the Google Play Store. Features of Box include the ability to share links to synced files to others and clients for devices such as PCs and mobile devices.

Readers should note that we did not investigate the 'OneCloud' feature of Box as its app store like functionality was not within the scope of our research.

For Box:
- The **private app storage path** is "/data/data/com.box.android"
- The **external storage data path** is "/sdcard/Android/data/com.box.android"

### 2.2.1 Examine App Files in Private Storage (Box)

Box stores a range of files in its private app directory including configuration files, user file previews/thumbnails and a specialized key-value pair database. These artefacts are discussed in detail below.

**[private app storage path]/files/leveldb[User ID]/**
- This directory contains the constituent files which describe the levelDB key-value pair database. This database is used heavily by Box to store data and metadata, and it is discussed further in the database examination section below.

**[private app storage path]/files/previews/**
- This directory contains "previews" of the files which have been cached by the Box app. These files can be thumbnails (for videos and images) or the actual files (e.g. PDFs). The files stored in this directory are unencrypted.
- Filenames in this directory generally use the format: preview_[file_ID]_1_[int]_.[file_extension].

**[private app storage path]/files/thumbnails/**
- This directory contains the cache of generated thumbnails used in Box. The files are stored in standard image file formats (e.g. PNG and JPEG). The filenames contain a range of information including the timestamp of thumbnail generation with millisecond resolution in UTC, the file ID, and dimension information (where applicable).
- In our experiments, we found thumbnails generated in this directory from internal app facilities (such as "OneCloud" which include "onecloudapp" in the filename), files uploaded by the user which include "file" in the filename and the user's avatar which includes "avatar" and the user's ID in the filename.

- The files in this directory directly relate to the entries in the 'imagecachedb' databases as discussed further below.

**[private app storage path]/shared_prefs/GLOBAL.xml**
- This XML formatted file contains a number of directives relating to app configuration.
- One directive is of particular interest; "storedLoggedInUsers" is a string value which stores a JSON formatted string containing:
    - id – The user's numerical ID.
    - userAuthToken – The current access token cached for the user.
    - userRefreshToken – The user's refresh token.
    - userName – The user's email address (used as a username for Box).

**[private app storage path]/shared_prefs/myPreference[User ID].xml**
- This XML formatted file contains a number of configuration directives for individual Box users.
- The first directive of interest is "com.box.android.encryptionKey", which is a string containing the 512-bit encryption key used by Box to encrypt files stored on external storage.
- The next directive of interest is "com.box.android.MoCoBoxUsers.userInfo", which is a JSON encoded string that contains:
    - "login" and "name" – The user's email address (used as a username for Box).
    - id – The user's numerical ID.
    - avatar_url – The unauthenticated URL used to retrieve the user's avatar image.
    - max_upload_size – The maximum permitted upload size represented in bytes in exponential notation
    - space_amount – The total quota available to the user represented in bytes in exponential notation.
    - space_used – The total space consumed by the user represented in bytes in exponential notation.

**[private app storage path]/shared_prefs/PREVIEW_SALTS[USER ID].xml**
- This XML formatted file contains the salts of each encrypted preview file stored on external storage. These salts are generated when the preview is stored.
    - Salts are stored as strings with the file ID (of the source file of the preview) as the 'name' and the salt as the value.

**[private app storage path]/shared_prefs/DOWNLOAD_SALTS[USER ID].xml**
- This XML formatted file contains the salts of each encrypted file cached on external storage. These salts are generated when the file is downloaded to external storage.
    - Salts are stored as strings with the file ID as the 'name' and the salt as the value.

### 2.2.2 Examine App Files on External Storage (Box)
Box uses external storage to maintain a cache of various previews and downloaded files.

The majority of files which Box stores on the external storage are encrypted using a format known as "Box Crypto". In order to decrypt these files, the appropriate salt must be found (based on the file's ID) in the "DOWNLOAD_SALTS[USER ID].xml" or the "PREVIEW_SALTS[USER ID].xml" files depending on whether a full file or a preview is being decrypted respectively. The encryption key is obtained from the "myPreference[USER ID].xml" file. By utilizing both of these sets of data, the cached file can then be decrypted by following these steps:

1. Firstly the app's encryption key is appended with a "_" followed by the file's ID.
2. This concatenated string is then passed through the SHA1 algorithm a total of 10 times to produce a new string.

3. The file can then be decrypted by passing the file, the newly generated string (as the key) and the salt obtained above into Bouncy Castle's AES CBC cipher (using PKCS5Padding) for decryption.

In addition, Box stores the following files which are of forensic interest:

**[external storage data path]/[User ID]/cache/dl_cache**
- This directory contains the cache of files which have been downloaded by the user in Box Crypto format.
- Files in this directory have a file name format of: [File ID]_[SHA1 of the original file] with no file extension.

**[external storage data path]/[User ID]/cache/dl_offline**
- Contained within this directory are files that the user has specifically chosen to have offline access to. These files are also stored in the Box Crypto format.
- Similarly, files in this directory have a file name format of: [File ID]_[SHA1 of the original file] with no file extension.

**[external storage data path]/[User ID]/cache/previews**
- This directory contains the previews of files which have been viewed by the user in the Box Crypto format. Depending on the format of the original file, these files may be the entire original file or scaled thumbnails (in the case of images).
- The files in this directory have a file name in the following format: preview_[File ID]_[int]_[int].[file extension].

### 2.2.3 Examine App Databases (Box)

The Box app utilizes SQLite databases located within the "**[private app storage path]**/databases" directory. It also makes use of a key-value pair database format known as levelDB, used by Box to store data and metadata. The databases (and items within these databases) of interest are presented in Table 4, Table 5 (both SQLite databases) and Table 6 (a levelDB format database).

| **[private app storage path]/databases/BoxSQLiteDB_[User ID]** ||
|---|---|
| "BoxEvent" Table (records of all actions performed by the app on the user's files) ||
| **Attribute** | **Description** |
| source_item_type | The item type of the source of the event (e.g. "file" or "folder"). |
| event_owner_id | The user ID of the user which triggered the event. |
| event_type | The type of action being triggered on the source item. Examples include:<br>• ITEM_COPY<br>• ITEM_PREVIEW<br>• ITEM_SHARED<br>• ITEM_CREATE<br>• ITEM_MOVE<br>• ITEM_DOWNLOAD<br>• ITEM_UPLOAD |
| source_item_id | The ID for the item which is being affected by the event. |
| created_at | A millisecond resolution timestamp representing the time at which the event was initiated. |
| user_dismissed | A Boolean value representing if the user has dismissed the notification in the Box UI. |
| parent_id | This field was blank in our experiments. |
| name | The event's name comprising the string "event_" and the event's ID. |
| modified_at | An integer which was 0 for all entries in our experiments. |
| size | A double which was 0.0 for all entries in our experiments. |
| id | The event's unique identifier. |

| "BoxFile" Table (metadata for files stored in Box) | |
|---|---|
| parent_id | The ID of the parent directory, 0 if root of the directory structure. |
| name | The filename as entered by the user including any extensions. |
| modified_at | The timestamp with millisecond resolution which represents the last modification to the file. |
| size | The file's size in bytes. |
| id | The file's ID. |
| "BoxFolder" Table (metadata for folders created in Box) | |
| parent_id | The ID of the parent directory, blank for the root of the directory structure. |
| name | The folder name as entered by the user. |
| modified_at | The timestamp with millisecond resolution which represents the last modification to the contents of the folder. |
| size | The size in bytes of all contents of the folder combined. |
| id | The folder ID, 0 for the root of the directory structure. |
| "BoxRecentFile" Table (metadata for recent files accessed in Box) | |
| item_id | The ID of the recently accessed file. |
| item_type | The item type, "file" in our experiments. |
| recent_item_id | The ID of the recently accessed file, identical to "item_id" in our experiments. |
| user_dismissed | A Boolean presumably representing if the user has dismissed the recent file from the list, "0" in our experiments. |
| timestamp | A millisecond resolution timestamp representing the time at which the file was accessed. |
| id | An ID consisting of the "item_type" and "recent_item_id" concatenated with an underscore. |
| "BoxComment" Table (metadata for comments made on files in Box) | |
| created_at | A human readable timestamp representing when the comment was made. |
| item_id | The item on which the comment was made. |
| item_type | The type of item on which the comment was made. |
| id | The comment ID as referenced in levelDB. |

*Table 4 "BoxSQLiteDB_[User ID]" Database File*

| [private app storage path]/databases/imagecachedb | |
|---|---|
| "files" Table (Stores thumbnail information for media) | |
| **Attribute** | **Description** |
| _id | The ID of this particular record in the table. |
| timestamp | A millisecond resolution timestamp representing when the image was first cached. |
| url | The "url" for this particular item in the format [object type]_[item id]_[int]_[dimension]. |
| image_filename | The filename of the thumbnail as stored within the "**[private app storage path]**/files/thumbnails" directory. |

*Table 5 "imagecachedb" Database File*

| [private app storage path]/files/leveldb[User ID]/ | |
|---|---|
| boxitem://comment/[comment ID] | |
| **Attribute** | **Description** |
| type | The type of entry i.e. "comment". |
| item | An array which contains the values "type" and "id". In our case, "type" was "file" and "id" was the file ID of the file to which the comment was added. |
| message | The comment text as entered by the user. |
| id | The comment ID. |
| created_by | An array which contains information about the user who created the comment. It comprises the values "type", "login", "name", and "id". In our case, the "type" was "user", the "login" and "name" were the user's email address and the "id" is the user's ID. |
| is_reply_comment | A Boolean representing if the comment is a reply. |

| | | |
|---|---|---|
| | created_at | A human readable timestamp representing when the comment was made. |
| colspan="2" | boxitem://file/[file ID] | |
| | type | The type of entry i.e. "file". |
| | parent | An array which contains information about the parent folder of this file. This array contains the "type" (folder), the name (folder's name), and folder id. |
| | permissions | An array containing an assortment of Boolean values relating to what actions the user is able to perform on this file. These values are:<br>• can_comment<br>• can_delete<br>• can_download<br>• can_preview<br>• can_rename<br>• can_set_share_access<br>• can_share<br>• can_upload |
| | sha1 | The SHA1 hash of the file. |
| | name | The full filename for this file including any extensions. |
| | size | The size of the file in bytes (in exponent form). |
| | id | The file's unique identifier. |
| | path_collection | An array containing the folders which make up the path to this particular file. For example, files located in the root directory have a single folder's (root) information listed within this array. See "parent" for this folder information. |
| | shared_link | This is an (optional, depending on whether the file has been shared) array containing information about this file's link sharing. Included within this array is another array called "permissions" (Booleans listing whether the shared file can be downloaded/previewed). Other values include:<br>• access – "open" if available to the public.<br>• url – the URL to the shared file.<br>• download_count – Integer value representing how many times the file has been downloaded.<br>• download_url – A static URL for the file.<br>• preview_count – Integer value representing how many times the shared file has been previewed.<br>• is_password_enabled – Boolean representing if the file has been password protected. |
| | comment_count | An integer representing the number of comments made on this file. |
| | content_created_at | A human readable timestamp representing when the file was created. |
| | content_modified_at | A human readable timestamp representing when the file was last modified. |
| | modified_by | An array containing information on the user who last modified the file (see "created_by" in boxitem://comment). |
| | owned_by | An array containing information on the owner of the file (see "created_by" in boxitem://comment). |
| colspan="2" | boxitem://folder/[folder ID] | |
| | type | The type of entry i.e. "folder". |
| | permissions | (see "permissions" in boxitem://file) |
| | name | The folder's name. |
| | size | The size of all the combined contents of the folder, represented in exponent form, in bytes. |
| | id | The folder's unique identifier. |
| | path_collection | (see "path_collection" in boxitem://file) |
| | has_collaborations | A Boolean value representing if the folder has any collaborators. |
| | modified_by | (see "modified_by" in boxitem://file) |
| | owned_by | (see "owned_by" in boxitem://file) |
| colspan="2" | boxitem://event/[event ID] | |
| | type | The type of entry, i.e. "event". |
| | source | The source file of the event. This is a "boxitem://item". |
| | event_type | The type of event that is being recorded. See "event_type" in Table 4. |
| | event_id | The unique identifier for the event. |

| created_by | (see "modified_by" in boxitem://file) |
| --- | --- |
| created_at | A human readable timestamp representing when the event was triggered. |

*Table 6 LevelDB File*

### 2.2.4 Examine and Analyze Accounts Data (Box)

After replacing the services.jar file within the "/system" partition and running our signed app, we found that Box does not store data using the AccountManager API on Android devices. This means that a Box account (pertaining to the currently logged in user) does not appear in the system Settings app under the list of accounts currently on the device and, therefore, no data relating to Box or the Box app was located using AccountManager.

## 2.3 ONEDRIVE

OneDrive (formerly known as SkyDrive) is another popular file-syncing storage service created by Microsoft. It has over 5 million downloads on the Google Play Store and is capable of interacting with the Microsoft Office suite of products.

Our findings are described in detail below. For OneDrive:
- The **private app storage path** is "/data/data/com.microsoft.skydrive"
- The **external storage data path** is "/sdcard/Android/data/com.microsoft.skydrive"

### 2.3.1 Examine App Files in Private Storage (OneDrive)

In our experiments, we found five XML format files within the "shared_prefs" directory inside the private app storage directory of the OneDrive app. The configuration directives in these files (approximately eight in total) all relate to the app configuration. These files contained no data which we considered to be of general interest to a forensic practitioner.

### 2.3.2 Examine App Files on External Storage (OneDrive)

OneDrive stores a cache of files downloaded by the user on the device's external storage. Cached files are located at the "[**external storage data path**]/cache" location. They follow the "SkyDriveCacheFile_[integer].cachedata" filename convention, where the integer identifies the file based upon its "id" in the "cached_files_metadata" table of the "cached_files_md.db" database – see Table 8. Cached files are stored on the external storage unmodified (hashes match for original files uploaded and files cached).

### 2.3.3 Examine App Databases (OneDrive)

OneDrive stores five SQLite format databases in its **[private app storage path]/databases** directory. We consider two of these databases, "cached_files_md.db" & "metadata" (see Table 7), to be of particular forensic interest and outline the contents of "auto_upload.db" and "manual_upload_db" in Table 9 and Table 10 respectively. In our experiment environment, "auto_uploaded_files_md.db" did not contain any data of general forensic interest.

| [private app storage path]/databases/metadata | |
| --- | --- |
| "items" Table (stores metadata for items in the users OneDrive) | |
| **Attribute** | **Description** |
| _id | An integer identifier for the record. |
| parentRid | The resource ID of the parent folder of this file/folder if applicable. For example, root does not have a parentRid, and items stored in the root have the parentRid 'root'. |
| ownerCid | The client ID of the owner of the file/folder as stored in account manager extras. |
| resourceId | The unique (string based) identifier for this resource. |
| parentId | The parent "_id" for this item. |
| downloadUrl | The URL used to download the item, where applicable. This URL requires authentication. |

| Attribute | Description |
|---|---|
| extension | The file extension where applicable. |
| lastAccess | Some items store a last accessed timestamp in millisecond resolution. |
| modifiedDateOnClient | A millisecond resolution timestamp representing the last modified date as reported in the OneDrive app. |
| creationDate | A millisecond resolution timestamp representing the date the item was added to OneDrive. |
| name | The friendly name of the item as set by the user. |
| ownerName | The first name and last name as entered by the user for their account. |
| sharingLevel | A string representation of the sharing enabled on this item (e.g. "Just Me" and "Public"). |
| size | The size of the item in bytes; in the case of a folder, the size of the sum of its contents. |
| size_text | A human readable representation of the size field (e.g. "100 KB"). |
| totalCount | The number of child items inside this item. |
| mimeType | The mime type of the item. |
| eTag | The eTag is a concatenation of "[resourceId].[version number]". We presume the final integer after the decimal is a version number as it is generally 0 for items which have not been modified and greater than 0 for those which have. |

*Table 7 OneDrive "metadata" Database*

| [private app storage path]/databases/cached_files_md.db | |
|---|---|
| "cached_files_metadata" Table (stores metadata for cached files) | |
| Attribute | Description |
| id | An integer identifier for the record. |
| cache_id | The item's "resourceId" from the metadata database concatenated with "_Download". |
| skydrive_url | The item's "downloadUrl" from the metadata database. |
| etag | The item's "eTag" from the metadata database. |
| last_access_time | The timestamp representing the last time the cached item was accessed. |
| file_size_bytes | The size of the cached item in bytes. |
| is_at_internal_storage | Blank in our experiments as files are stored on the external storage. Presumably 1 where files are stored on internal storage. |

*Table 8 "cached_files_md.db" Database File*

| [private app storage path]/databases/auto_upload.db | |
|---|---|
| "queue" Table (queueing information for files waiting to be uploaded) | |
| Attribute | Description |
| _id | An integer identifier for the record. |
| creationDate | A timestamp representing the date the file was added to the queue for uploading. |
| fileName | A concatenation of the date (in YYYYMMDD format), time (in 24hr time HHMMSS) and "Android.jpg". |
| fileNameOriginal | The original filename set by the user/camera app. |
| filePath | The path on the device to the file to be uploaded. |
| fileSize | The file size in bytes. |
| loadingProgress | Presumably the number of bytes which have been transferred (0 in our experiments). |

*Table 9 "auto_upload.db" Database File*

| [private app storage path]/databases/manual_upload_db | |
|---|---|
| "queue" Table (contains information about the user's files which were manually uploaded) | |
| Attribute | Description |
| _id | An integer identifier for the record. |
| fileName | A concatenation of the date (in YYYYMMDD format), time (in 24hr |

| | time HHMMSS) and Android.jpg. |
|---|---|
| filePath | The path on the device to the file which is to be uploaded. |
| fileSize | The file size in bytes. |
| folderOwnerCid | The client ID of the owner of the folder the file is being uploaded to. |
| folderResourceId | The "resourceId" of the folder the item is being uploaded to. |
| loadingProgress | Presumably the number of bytes which have been transferred (the size of the file in bytes in our experiments). |

*Table 10 "manual_upload_db" Databases File*

### 2.3.4 Examine and Analyze Accounts Data (OneDrive)

We installed our appropriately signed APK into an Android system that had been updated with our modified services.jar file and found that OneDrive stores a significant amount of data within the AccountManager service on the Android device. This may explain the lack of accounts data which was of interest within the OneDrive app's private directory. Several items of interest were found within the AccountManager API (see Table 11):

| AccountManager API Calls | |
|---|---|
| **Method** | **Description** |
| getPassword() | Returns the OneDrive app's current refresh token. |
| getAuthToken() | Returns a large amount of data including the current refresh token, access token, scope, account type, user ID and access token expiry timestamp. |

*Table 11 AccountManager (OneDrive)*

## 2.4 ownCloud

ownCloud is a popular open source alternative to the above file syncing apps. Users are able to freely host their own private ownCloud servers and set up a private file sync service. The Android app itself is a paid app on the Google Play Store but as the app is open source, it can simply be built from the publicly available code.

ownCloud utilizes the following paths:
- The **private app storage path** is "/data/data/com.owncloud.android"
- The **external storage data path** is "/sdcard/owncloud"

### 2.4.1 Examine App Files in Private Storage (ownCloud)

ownCloud stores its configuration in a single shared_prefs XML file named "com.owncloud.android_preferences.xml", the configuration directives in this file are discussed below:

**[private app storage path]/shared_prefs/com.owncloud.android_preferences.xml**
- "instant_upload_on_wifi" – This Boolean value specifies whether instant upload for pictures only occurs when connected to Wi-Fi.
- "instant_uploading" – This Boolean value specifies whether instant upload for pictures is enabled.
- "select_oc_account" – This string value represents the ownCloud account (username@server) which should be selected by the app upon launch.
- "set_pincode" – This Boolean value specifies whether an "app PIN" (a four digit PIN which must be entered to use the app on launch) is enabled.
- "PrefPinCode[1-4]" – These string values represent the individual integers of the 4 digit PIN code if enabled.

Excluding the database files discussed below, no other files of interest were found in this app's private storage directories.

### 2.4.2 Examine App Files on External Storage (ownCloud)

ownCloud stores cached and downloaded (favorite) files in a mirror of the server directory structure (for the parents of downloaded files) on external storage at the following path [**external storage data path**]/[username@server]. Files are stored in this directory structure unmodified.

### 2.4.3 Examine App Databases (ownCloud)

The ownCloud Android app stores two SQLite format database files within its [**private app storage path**]/databases directory. Both of these files contain data that may be of general forensic interest and as such are outlined in Table 12 and Table 13.

| [private app storage path]/databases/filelist | |
|---|---|
| "filelist" Table (contains information about all of the files/folders of all the users that have authenticated in the ownCloud app) | |
| **Attribute** | **Description** |
| _id | An integer identifier for the record. |
| filename | The file name of the file including any extensions. |
| path | The path to the file from the root directory of the ownCloud server. |
| parent | The "_id" of the parent directory to this file/folder. |
| modified | A millisecond resolution timestamp representing when the cached file was last modified. |
| content_type | The mime type of a file or "DIR" for a folder. |
| Media_path | The full path to the file on the Android device's external storage. |
| File_owner | The owner of the file in the format of "username@server". |
| Last_sync_date | A millisecond resolution timestamp representing when the file was lasted synced with the ownCloud server. |
| keep_in_sync | A Boolean value representing whether a file should be kept in sync. This function is manually enabled by the user. |
| last_sync_date_for_data | A millisecond resolution timestamp representing the last time a file was downloaded by the user. |
| modified_at_last_sync_date_for_data | A millisecond resolution timestamp representing the file modified date on the ownCloud server at last sync. |
| share_by_link | A Boolean value representing if a file has been shared. |
| etag | A unique identifier (on a per account basis) for each item used for caching purposes. Only folders had an "etag" recorded in our experiments. It is generated from the server's "eTag" database record. |
| "ocshares" Table (metadata for the shared files supported by the ownCloud mobile app) | |
| _id | An integer identifier for the record. |
| file_source & item_source | The "_id" from the filelist table for the file being shared |
| shate_with *[sic]* | The password required to access the share URL (blank if not set). This password is stored as a Blowfish hash. |
| path | The path to the file from the ownCloud root directory for the user. |
| shared_date | A second resolution timestamp indicating when the share was created. |
| expiration_date | A second resolution timestamp indicating when the share will/has expired. |
| token | The token used to access the file in the share URL, e.g. "http://[ocserver]/owncloud/public.php?service=files&t=[token]". |
| is_directory | A Boolean value representing whether the item is a directory. |
| owner_share | The owner of the shared item in the format of "username@server". |

*Table 12 "filelist" Database File*

| [private app storage path]/databases/ownCloud | |
|---|---|
| "instant_upload" Table (contains information about pictures selected for auto upload, if this feature is enabled. Records are deleted once the upload is successful.) | |
| **Attribute** | **Description** |
| _id | An integer identifier for the record. |
| path | The path on local storage to the file which is to be uploaded. |
| account | The ownCloud account for the user the images will be uploaded to in the format of "username@server". |

*Table 13 "ownCloud" Database File*

### 2.4.4 Examine and Analyze Accounts Data (ownCloud)

By utilizing our installed app that is able to bypass system signature checking, we were able to obtain the data that ownCloud stored within the AccountManager API. We found that ownCloud stores the user's password in plaintext when using the "getPassword()" method of the AccountManager API.

## 2.5 EVERNOTE

Evernote is one of the most popular note taking apps in the Google Play Store, with over 50 million downloads. Its features include the ability to recognize text from handwritten notes (OCR) and cross-platform note syncing capabilities.

For Evernote:
- The **private app storage path** is "/data/data/com.evernote"
- The **external storage data path** is "/sdcard/Android/data/com.evernote"

### 2.5.1 Examine App Files in Private Storage (Evernote)

Evernote creates a number of files in its private storage directories. Within the "shared_prefs" directory Evernote stores approximately seventeen XML configuration files, of which we have selected four which we found to be of forensic interest.

**[private app storage path]/shared_prefs/[User ID].pref.xml**
- "userid" – The unique identifier for the user (an integer in our case).
- "username" – The user's Evernote username.
- "encrypted_authtoken" – A Base64 encoded string. Presumably an encrypted copy of the authentication token. This token can be decrypted using Evernote's "com.evernote.util" classes. These classes are heavily obfuscated.
- "default_notebook" – The GUID of the default notebook.
- "AcctInfoWebPrefixUrl" – The URL used as part of account authentication.
- "email" – The user's email address.
- "LAST_USER_OBJECT_SYNC_TIME" – A millisecond resolution timestamp which represents the last sync time.
- "LAST_DB_FILEPATH" – A string listing the file path to the Evernote database. Notably this database was stored on external storage in our experiments.
- "Last_server_acc_info_timestamp" – Timestamp of last successful login to the Evernote servers (i.e. last online session).
- "AcctInfoNoteStoreUrl" – The URL used by the app for x-thrift communication.

**[private app storage path]/shared_prefs/[User ID]_counts.pref.xml**
- The counts configuration file maintains a listing of the number of objects in seven categories, namely: "places", "notes", "sktiches", "tags", "notebooks", "snotes" and "linked notebooks".

**[private app storage path]/shared_prefs/[User ID]_sync_state.pref.xml**
- The sync state preference file stores a number of directives. We found the "SYNC_STATUS_MSG" to be of use as it describes the last sync information in a human readable format (e.g. "Last sync: 1 Jan 12:00 pm").

**[private app storage path]/shared_prefs/ com.evernote_preferences.xml**
- "PREF_USERID_LIST" – The user ID of the logged in user. Presumably if more than one user could be logged in on the device, they would be listed here.
- "PREF_ACTIVE_USERID" – The user ID of the currently active user.
- "last_viewed_notes" – The GUID of the note last viewed by the user.

**[private app storage path]/files/.logs/log_file.txt**

- This is a time-stamped log file generated by the Evernote app that contains events. The events recorded include files being opened, stored and user login events. The log contains the user ID of the user to which the recorded event pertains.

**[private app storage path]/files/.usercache/user.dat**
- This is a serialized file created by the Evernote app.

### 2.5.2 Examine App Files on External Storage (Evernote)

Evernote stores a number of items on the device's external storage. These files are:

**[external storage data path]/files/user-[User ID]/mapthumbdb/thumbnails_data_1.dat**
- This file contains a cache of thumbnails generated in the Evernote app. We were able to recover a JPEG format image from this file using header analysis.

**[external storage data path]/files/notes/[first three characters of GUID]/[note GUID]/**
- This directory contains a note as identified by the GUID and the constituent files (where relevant) such as image files.
    - The main content of the note is stored in a "content.enml" file, which is a form of human readable XML encoding used by Evernote.
    - The content of the note may also be represented in HTML with a filename starting with "note" with the "html" extension.
    - Images and other file content are represented using GUIDs with the "dat" extension. Metadata for each of these files is available in the "resources" table of the Evernote database.

Evernote also stores its database files on external storage, this is discussed further below.

### 2.5.3 Examine App Databases (Evernote)

In our experiments, we found that Evernote does not utilize the databases folder within its private app storage to store its app's databases. Instead Evernote stores its database in an unencrypted file on the device's external storage. The database and its contents are described in Table 14.

| [external storage data path]/files/user-[User ID] | |
|---|---|
| "guid_updates" Table (Presumably a list of GUIDs which have been updated) | |
| **Attribute** | **Description** |
| new_guid | The newly generated globally unique identifier (GUID). |
| old_guid | The previous GUID. |
| "note_tag" Table (a relationship table between tags and notes) | |
| note_guid | The note's GUID from the "notes" table. |
| tag_guid | The tag's GUID from the "tags_table" table. |
| "notebooks" Table (a table containing metadata relating to the user's notebooks) | |
| guid | The GUID of the notebook. |
| name | The friendly name of the notebook as set by the user. |
| published | A Boolean value representing if the notebook has been shared. |
| "notes" Table (a table containing metadata relating to the user's notes) | |
| guid | The GUID of the note. |
| notebook_guid | The GUID of the notebook parent for this note. |
| title | The title of the note. |
| content_length | The size of the note content (not including resources such as images) in bytes. |
| content_hash | A binary hash of unknown type. |
| created | A timestamp with millisecond resolution representing when the note was created. |
| updated | A timestamp with millisecond resolution representing when the note |
| | was last modified. |
| deleted | A timestamp with millisecond resolution representing when the note was deleted (where applicable otherwise 0). |
| is_active | A Boolean value representing the file's deleted state. |
| cached | A Boolean value representing whether the file has been cached on the device. |
| "city", "state", "country", "latitude", "longitude", "altitude" | The location data for the note, if enabled. |
| source | The source of the note (e.g. "mobile.android"). |
| source_url | The source URL for the note where applicable (e.g. "http://evernote.com/"). |
| note_share_date | A timestamp with millisecond resolution representing when note sharing was enabled. |
| note_share_key | The key component of the sharing URL which follows the following format: "[AcctInfoWebPrefixUrl]/sh/[note GUID]/[note_share_key]". |
| task_date | A timestamp with millisecond resolution representing when the note was created for some notes. |
| task_complete_date | A timestamp with millisecond resolution representing when the note was marked as "Done". |
| task_due_date | A timestamp with millisecond resolution representing when the task is set to be due by the user. |
| "resources" Table (a table containing metadata relating to resource objects used in notes) | |
| guid | The resource GUID. |
| note_guid | The GUID of the parent note of the resource. |
| mime | The mime type of the resource. |
| width | The width of an image. |
| height | The height of an image. |
| hash | A binary hash of unknown type. Used as the file name of the resource image on the file system (the binary hash is converted to hexadecimal). |
| cached | A Boolean value representing whether the resource has been cached on the device. |
| length | The size of the resource in bytes. |
| has_recognition | A Boolean value presumably representing whether the resource has undergone character recognition. |
| timestamp | A timestamp with millisecond resolution for the resource (e.g. when the image was taken). |
| filename | The resource filename (where applicable). |
| reco_cached | A Boolean value presumably representing whether the resource character recognition is cached. |
| ink_signature | A JSON encoded string which presumably includes ink related metadata including height, width and a GUID. |
| "search_history" Table (a table containing metadata relating to search history) | |
| query | The search queries entered by the user in the Evernote app. |
| updated | A timestamp with millisecond resolution representing when the search using the keywords in "query" was last performed. |
| "search_index_content" Table (a table containing metadata relating to searches) | |
| c0note_guid | The GUID of the note which will be returned from matching searches. |
| c1content_id | The resource GUID (for images) or the file type. |
| c3keywords | A string of searchable content derived from the note and resources. |
| "snippets_table" Table (a table containing "snippets" of information about each note) | |
| note_guid | The GUID of the note to which the snippet relates. |
| mime_type | The mime type of the snippet where relevant. |
| res_count | Presumably the number of resources used in the note. |
| snippet | The first 193 characters of the note text, where applicable. |
| "tags_table" Table (a table containing metadata relating to tags) | |
| guid | The GUID of the tag. |
| name | The name of the tag as entered by the user. |

*Table 14 "user-[User ID]" Database File*

### 2.5.4 Examine and Analyze Accounts Data (Evernote)

The Evernote Android app utilizes the AccountManager API to store some account-related data. However, we found that Evernote does not store a "password" for the logged in account as accessible via the AccountManager API's "getPassword()" method. AccountManager's "getAuthTokens()" method also did not return any data.

## 2.6 ONENOTE

Microsoft's OneNote is a note taking app with cloud syncing capabilities. It is also (like Evernote) a cross-platform app. As OneNote is a Microsoft app, it utilizes package names similar to OneDrive. OneNote has the following paths:
- The **private app storage path** is "/data/data/com.microsoft.office.onenote"
- The **external storage data path** is "/sdcard/Android/data/com.microsoft.office.onenote"

Our findings for OneNote are described in the following sections.

### 2.6.1 Examine App Files in Private Storage (OneNote)

OneNote has two configuration files in its "shared_prefs" directory, "com.microsoft.office.onenote.eula2.xml" and "com.microsoft.office.onenote_preferences.xml". The former file stores EULA information which we do not consider to be of forensic interest, and the latter file stores general configuration for the OneNote app. We highlight the directives of interest in this file below.

**[private app storage path]/shared_prefs/com.microsoft.office.onenote_preferences.xml**
- "KEY_RESUME_VIEW_ID" – This string stores the GUID of the last notebook viewed by the user.
- "DEFAULT_LIVE_ID" – This string is used as the identifier for the user's Live ID.

Within its "files" subdirectory, OneNote stores a number of files. Within the root of this directory is a "registry.xml" which stores an XML encoded registry file.

**[private app storage path]/files/registry.xml**
- "SQL DB Path" – The value of this key represents the path on the local device to the SQL database directory for OneNote's file store.

If a practitioner navigates to the path listed in the file, they should locate a "File Store" directory with a number of subdirectories uniquely identifying the OneNote user. An example of the path and files stored in this location is as follows:

**[private app storage path]/files/Microsoft/Office Mobile/SPM Data/File Store/1000/https/d.docs.live.net/[ DEFAULT_LIVE_ID]**
- "{[section GUID]}.one" – Cached versions of the OneNote sections which have been accessed/created by the user. A notebook comprises a collection of sections (as defined in the OneNote database). The files are in the standard "one" OneNote format and can be read using the OneNote PC application.

### 2.6.2 Examine App Files on External Storage (OneNote)

OneNote creates the "[external storage data path]/files" path on the device's external storage. However, in our experiments, we were unable to locate any files in this directory.

### 2.6.3 Examine App Databases (OneNote)

OneNote does not create a "databases" directory within the private app storage path, but rather stores databases in various locations throughout private app storage subdirectories. In total, we located two "sdf" format databases of interest, namely: "[private app storage path]/files/Microsoft/Office Mobile/SPM Data/SPSQLStore.sdf" and "[private app storage path]/files/OneNote/hierarchy.sdf". The former file contains metadata relating to OneNote's constituent files, and the latter contains metadata relating to the notes stored in the app. The SDF format file headers identify the files as "SQLite format 3". These databases are outlined in Table 15 and Table 16.

| [private app storage path]/files/Microsoft/Office Mobile/SPM Data/SPSQLStore.sdf | |
|---|---|
| "SPMCConfigData" Table (contains general information about the account) | |
| **Attribute** | **Description** |
| FieldName | This includes fields such as:<br>• NewDefaultNotebookName<br>• SkyDriveRootDavUrl<br>• SkyDriveSignedInUser |
| FieldValue | This contains the values for the above "FieldName" fields. |
| "SPMCItems" Table (metadata regarding the notebooks and section files) | |
| ObjectID | The item's GUID. |
| ListID | The GUID of the list relating to the item (relates to the SPMCLists table). |
| FolderID | The GUID of the parent folder (where applicable). |
| SiteID | The GUID of the related site (relates to the SPMCLists table). |
| ContentType | The item type (e.g. "Folder" or "Document"). |
| Created | A human readable creation time of the object, in the format YYYYMMDD HH:MM:SS. |
| Modified | A human readable modified time for the object, in the format YYYYMMDD HH:MM:SS. |
| FileDirRef | The parent directory path on the server for the object. |
| ProgId | The ID of the program used to open the file (e.g. "OneNote.Notebook"). |
| ServerUrl | The URL (when appended to the "SkyDriveRootDavUrl") which is used to access the item. |
| LinkFileName | The item name including extension (where applicable). |
| EncodedAbsUrl | A "URL Encoded" complete URL for accessing the item. |
| FileType | The file extension (where applicable). |
| Etag | This field mirrors the modified date for documents, presumably used for cache management. |
| FileSize | The item size in bytes (where applicable). |
| LevelDescription | The item's sharing information (e.g. "Shared with: Just me" – where applicable). |
| "SPMCObjects" Table (sync information for the items in OneNote) | |
| ObjectID | The object's GUID. |
| LastSyncTime | A human readable timestamp representing the time of last attempted object sync, in the format YYYYMMDD HH:MM:SS. |
| Deleted | A Boolean representing if the object has been deleted. |
| IsOnServer | A Boolean representing if the object is located on the server. |
| LastSuccessSyncTime | A human readable timestamp representing the time of last successful object sync, in the format YYYYMMDD HH:MM:SS. |
| DisplayTitle | The objects name (e.g. "Quick Notes.one") |
| UrlString | A serialized string containing parts of the information in this record. |
| ResId | The GUID or OneDrive File ID depending on the type of object. |
| CreatedTime | A human readable timestamp representing the time of creation of the object, in the format YYYYMMDD HH:MM:SS. |

*Table 15 "SPSQLStore.sdf" Database File*

| [private app storage path]/files/OneNote/hierarchy.sdf | |
|---|---|
| "OnmConfigData" Table (contains information about the configuration of notebooks) | |
| **Attribute** | **Description** |
| FieldName | This includes fields such as:<br>• UnfiledSectionID<br>• SkyDriveDefaultNotebookID<br>• LastSuccessfulUpdateNBListTime |
| FieldValue | This field contains the values for the above "FieldName" fields. |
| "OnmNotebookContent" Table (metadata for each of the users notebooks) | |

| ObjectID | The content's GUID. |
|---|---|
| ParentID | The GUID of the parent of this content. If this content is a notebook, then its "ParentID" is itself. If it is a section, then its "ParentID" is the notebook in which it resides. |
| ParentNoteBookID | This field mirrors "ParentID" in our experiments. |
| Name | The name of the content (e.g. "Quick Notes", "First Name's Notebook"). |
| DisplayName | This field mirrors "Name" in our experiments. |
| LastAccessTime | A human readable timestamps that was always "19000101 12:00:00" in our experiments. |
| LastModifiedTime | A human readable timestamp representing when the content was last modified, in the format YYYYMMDD HH:MM:SS. |
| "OnmSectionContent" Table (metadata for section content i.e. notes) | |
| ObjectID | The section content's GUID. |
| JotID | A GUID which was not found in other sections of the database or configuration. |
| ParentID | The ID of the parent object, and in our records these IDs represented sections (see OnmNotebookContent table). |
| Name | The name of the content (e.g. the note name set by the user). |
| LastAccessTime | A human readable timestamp representing when the content was last accessed, in the format YYYYMMDD HH:MM:SS. |
| LastModifiedTime | A human readable timestamp representing when the content was last modified, in the format YYYYMMDD HH:MM:SS. |
| Viewed | A Boolean value representing if the note has been viewed. |
| CreationTime | A human readable timestamp representing when the content was created, in the format YYYYMMDD HH:MM:SS. |

*Table 16 "hierarchy.sdf" Database File*

### 2.6.4 Examine and Analyze Accounts Data (OneNote)

Unlike many Android apps, Microsoft's OneNote stores a significant amount of information within the AccountManager API. Using AccountManager's "getPassword()" method, we found that OneNote stores a string that was not the user's login password . This "password" is used as part of an encryption cipher for which it is the encryption key. This is used in conjunction with the additional information that OneNote stores (accessible via AccountManager's "getUserData()" method).

Within the AccountManager "UserData" key-value store, OneNote stores two values of interest. Firstly, it stores the Live ID identifier of the logged in user. It also stores a string that contains an XML file. The contents of this string include:

- "_SEED" – The seed used to decrypt the refresh token.
- "LIVE_ID_FRIENDLY_NAME" – The user's full name (i.e. first name and last name).
- "_PASSWORD" – The user's encrypted refresh token.
- "_ID" – The user's Live ID identifier.
- "_LAST_MODIFIED" – A millisecond resolution timestamp representing when the XML string was last modified.

By using the above "_SEED" and "_PASSWORD" (decoded from Base64) and the "password" from the AccountManager "getPassword()" method, the unencrypted refresh token can be obtained. OneNote uses the AES standard for encryption and decryption, and, as such, the above values can simply be placed into an AES decryption method without any further modifications.

This concludes our analysis of the six cloud apps, which directly communicate with servers.

For completeness, we chose to perform evidence collection and analysis on an app which utilizes other (cloud) apps in order to sync its files.

## 2.7 UNIVERSAL PASSWORD MANAGER

Universal Password Manager or "UPM" is a password manager for Android devices with over 100,000 downloads. It does not store any details (such as usernames and passwords) on its own servers but rather in an encrypted database file stored on the user's Android device. It can utilize cloud storage services such as Dropbox, in order to allow the user to sync and backup the encrypted database amongst all their devices.

We chose an app that does not directly utilize its own cloud servers in order to demonstrate that our technique is applicable to more than just cloud-based apps. In principle, our process is capable of acquiring data from most Android apps.

Apart from the encrypted database, which is stored on the external storage at the appropriate location if the user opts to use a cloud syncing service, UPM does not store any additional files on external storage. This file should be stored in the cloud cache location such as within the external storage path of Dropbox, if cloud storage was utilized.

- UPM's **private app storage path** is: "/data/data/com.u17od.upm"

### 2.7.1 Examine App Files in Private Storage (UPM)

UPM stores several files of interest inside its private app storage path. Within the "shared_prefs" directory are two XML files of interest.

**[private app storage path]/shared_prefs/UPMPrefs.xml**
- "sync.method" – The value of this string represents the method of syncing the user has chosen for the UPM encrypted database. In our case, as we had chosen Dropbox as the syncing method, the value was "dropbox".

**[private app storage path] /shared_prefs/DROPBOX_PREFS.xml**
- "DROPBOX_SECRET" and "DROPBOX_KEY" – The values of these strings are used by Dropbox to authenticate the UPM app to allow it to sync the user's encrypted database. These strings are assigned by Dropbox to UPM when the user selects the sync method as Dropbox.

**[private app storage path]/files/upm.db**
- This file is the entirety of the encrypted UPM database containing all data the user has written into the UPM app. It is encrypted with a cipher known as "PBEWithSHA256And256BitAES-CBC-BC" and uses the user's selected password as a key. The salt for this cipher is stored in the encrypted database at positions 3 (with 0 being the first value) to 10 inclusively when read with a hex editor, for a total of 8 characters. With this knowledge in mind, it is possible to decrypt the database via brute force methods.

### 2.7.2 Examine App Files on External Storage (UPM)

UPM does not ordinarily utilize the device's external storage. However, if the user were to link the encrypted UPM database with a cloud storage syncing service (such as Dropbox), then the UPM encrypted database will be stored in that app's appropriate location for uploaded files.

### 2.7.3 Examine App Databases (UPM)

The UPM Android app does not use SQLite databases in order to store metadata as other apps have. This means the app does not create a "[private app storage path]/databases" directory.

### 2.7.4 Examine and Analyze Accounts Data (UPM)

UPM does not create an account in the AccountManager API for its Android app. As such, there is no UPM "account" on the device.

Based on these results, we moved to further analyze the seven apps we selected in order to determine if additional information could be obtained from them.

## 2.8 FURTHER APP ANALYSIS

The apps were further analyzed below in order to understand the underlying structure of the app, and to allow a forensic practitioner to use the information obtained in order to access a suspect user's files (which may not be stored on the device) without requiring modifications to the original device. At the end of each app's section, we give a verdict on whether (based on our findings) the practitioner would have enough information in order to fully authenticate as the user and access all files on the service's servers.

### 2.8.1 Dropbox Analysis

We collected a HPROF memory heap for the Android Dropbox app from our Android x86 virtual machine and then proceeded to analyze it using the Eclipse Memory Analyzer application.

Initial analysis of the Dominator Tree (sorted by Shallow Heap size) did not result in any significant findings. As such, we moved onto analysis using Object Query Language (OQL) statements. OQL searches were undertaken to locate URLs (querying for String objects which contain "https" or "dropbox.com"), authentication headers (querying for String objects which contain "Authorization") and user identification information (querying for String objects which contain email address attributes such as "@" as well as for strings such as username and password).

Using the OQL method, we were successful in locating the URLs that the Dropbox app uses for authentication and file retrieval. We were also successful in locating the authentication parameters required to use token authentication with Dropbox. The username of the logged in user was also located by searching for email addresses.

From the memory analysis of the Dropbox app alone, we were able to locate almost all of the information we propose would be required in order to authenticate as the user. This includes the authentication URL and the format of the header parameters for this URL. The initial analysis of the Dropbox app in Section 2.1 gave us the OAuth consumer key, the OAuth token and the second half of the OAuth signature (see Section 2.1.1). We were able to discern the missing string by referring to the Dropbox developer's guide [2].

As the value for the first half of the OAuth signature was not stored on the device (in our experiments), we concluded that this string must be statically assigned in the Dropbox app. Dropbox heavily obfuscates its code in order to discourage reverse engineering of its app. This means merely searching the decompiled source code for another static string used by Dropbox would not return any results (e.g. the OAuth consumer key is also generated statically). Certain strings cannot be fully obfuscated in the source code, such as URLs. Dropbox attempts to obscure these strings by building the URL as segments (e.g. "https://" + "dropbox.com") instead of having a single string object containing this information. By following these function calls through the code, we were able to locate a function that, using a static array of integers and a static seemingly random string of characters and integers, generated the OAuth consumer key and first half of the OAuth signature.

By collating all of the above information, a practitioner should be capable of authenticating as the user and accessing all files that the user had stored on the Dropbox servers.

### 2.8.2 Box Analysis

We were able to obtain a HPROF memory dump for the Box app. A cursory analysis of the Dominator Tree did not reveal any information of interest and so we proceeded to filter the Tree.

By narrowing down the memory data to only the data within the class path of the Box app ("com.box.android"), we were successful in locating two key pieces of information, namely: the "mClientId" used to identify the user of the app on the Box server and the "mClientSecret" which was used to verify this particular client (app) on the server. These were both stored as strings in an instance of the "com.box.android.boxclient.BoxSdkClient" class. These strings are important as they are part of the URL used to obtain a new OAuth access token. In order to further our analysis, we utilized OQL.

We used OQL statements to obtain all URLs contained within the memory of the Box app. This resulted in two URLs of interest, namely: A URL for displaying all items located within the user's root directory and a URL for generating a new access token for the user which requires the client id, client secret and refresh token to be transmitted as a POST request method. We were then able to determine that the client id and client secret must, therefore, be statically defined by Box. We decompiled the app and were able to locate these strings within the app's string resources file.

Our initial analysis of the Box app above (see Section 2.2) showed that the refresh token, the final piece of information we required to generate a valid access token, was stored in the Box app's private storage. We noted that access tokens expire after 60 minutes, making it unlikely that the access token stored within the app's private storage would be valid upon examination of the device by a practitioner. This meant that the refresh token and its URL found above would be useful in practice. We found that the refresh token expires after 60 days, and to generate a new access token, we only required a valid refresh token and the statically defined client id and client secret. Our analysis shows that we already have all of this information and, thus, should be able to obtain a valid access token given that the refresh token is still valid.

This access token could then be used with the URLs located for accessing all of a user's files. The same access token could be used (in the authorization header) to download any of the user's files. We do not believe that any additional information is required to access the user's files. Once again, we propose that access to all of the user's files would be attainable, following our analysis of the app.

### 2.8.3 OneDrive Analysis

Using the Eclipse Memory Analyzer tool, we were able to narrow down the entirety of the memory data (which includes Java and Android class objects) to only those within the app's class path ("com.microsoft.skydrive"). A simple search of this data resulted in the retrieval of the static "PRODUCT_ID" value in both integer and hexadecimal formats stored within an instance of the "com.microsoft.live.authorization.TokenRequest" class. This value (in hexadecimal format) is used by OneDrive to identify the app that is communicating with the server (known as the "client_id"). We did not find any other data of interest in the sorted and filtered Dominator Tree. As a result, we continued onto analysis using OQL statements.

Searching for all "https" URLs within the memory heap resulted in several key fragments of information. We were able to identify the URLs used by OneDrive to authenticate the user and obtain a new refresh token as well as the URLs to view all user files/folders and the URL to obtain a specific item within the user's synced storage. More importantly, we found the authentication header format used to authenticate the above URLs.

By searching the heap for terms such as "refresh" and "token", we were also able to obtain the JSON encoded replies from the server for the URLs used to obtain a new access token and refresh token. Contained within these JSON encoded replies were:

- **expires_at** – The timestamp in millisecond resolution of when the refresh/access token expires.
- **user_id** – A string generated when the OneDrive account was first created by the Microsoft Live server. It is used to identify the user on the server side.
- **refresh_token** – A new refresh token which can be used to obtain a new refresh and access token.
- **scope** – The scope (limitations) of the access token received, which is also used when obtaining refresh tokens and access tokens.

- **access_token** – A new valid access token which can be used to authorize the user to access their files and folders (with one of the above obtained URLs).

Based on this information, we now knew what data was required to authenticate as the user. The URL to obtain a new access token required a valid refresh token, a client id and a scope. From the memory heap, we already had the static client id of the OneDrive app. From the initial analysis of OneDrive in Section 2.3, we also had the refresh token and scope (both contained within the AccountManager API, see Section 2.3.4). Our analysis suggests that this is all the information required to obtain a new valid refresh token and an access token valid for 24 hours.

By utilizing this access token via the URLs used to list all user items, and then selecting an item from there to use the file downloading URL, we propose that a user's files could be obtained. The only parameter required in these URLs is the valid access token. It is, therefore, possible (with OneDrive) to obtain a user's files using only a valid refresh token (and, in turn, using the returned access token). All other values were static in our experiments.

### 2.8.4 ownCloud Analysis

After obtaining the HPROF memory heaps, we proceeded to analyze the Dominator Tree. The tree was filtered to exclude all references to non-ownCloud related classes. We noted that ownCloud keeps all user files that have been cached within the memory, including full paths to these files on the external storage. This information was stored under the class "OwnCloudFileObserver". Another class, "AccountUtils" stored the CARDDAV, ODAV, STATUS and WEBDAV paths on the ownCloud server.

We then proceeded to execute OQL queries on ownCloud's memory heap in order to further our examination of the app. By searching for all URLs within the memory heap, we were able to quickly locate the URLs used for authenticating with the ownCloud server, along with instances when the ownCloud app had accessed various locations within the server. Surprisingly, we were able to locate and obtain the user's username and password in plain text, which was used to authenticate the user into the ownCloud server. These details were stored in the Apache Commons "httpclient" class as opposed to residing within ownCloud's package. Unlikely the previously analyzed apps, ownCloud does not appear to use access and refresh tokens to authenticate the user but rather relies on sending the server the username and password each time authentication is required.

Based on the above memory analysis, we ascertained that ownCloud must store information such as the user's username and password on the device as it does not seem to utilize security tokens. We had found in Section 2.4.4 that ownCloud actually stores this information in the AccountManager API. By using the server authentication URL obtained as part of ownCloud's shared preferences file (see Section 2.4.1), it was possible to log in as the user (using their username and password) and gain access to all of their files.

### 2.8.5 Evernote Analysis

We obtained the HPROF memory heap of the Evernote Android app from our Android virtual machine. Firstly, we inspected the Dominator Tree of the memory heap in order to find artifacts of interest. By filtering the tree to show objects associated with the Evernote app ("com.evernote"), we were able to locate the (heavily obfuscated) URLs for authentication and for accessing user files. This was stored in the "com.evernote.client.w" class. Furthermore, this instance of the above class also contained the user's numerical ID, web browser agent and the parameters of the x-thrift format binary file that Evernote uses for authentication. These parameters include the user's authentication token.

In order to further our research, we proceeded to execute some OQL queries. Firstly, we searched for all URLs stored within the memory heap. The resulting URLs of interest were once again the URLs for authentication and accessing user notes. By searching for email addresses in the memory, we were able to obtain the email address of the Google account on the device, the email address the user used to sign up and log into Evernote and the Evernote specific email address.

We noted that the authentication URL and URLs for accessing files required an "x-thrift" format binary file to be sent in the body of the HTTP request. The contents of this binary file included the authentication token, although a significant amount of this file is in a binary format. As Evernote was heavily obfuscated, a decompilation of the code did not reveal data of significant interest.

Further analysis would be required in order to understand the inner workings of the Evernote app, to decrypt the encrypted token found in Section 2.5.1 and to understand the "x-thrift" format used by Evernote's server authentication.

### 2.8.6 OneNote Analysis

A cursory analysis of OneNote's HPROF file revealed that OneNote stores within its memory the full URL for authenticating with the Microsoft Live servers. Included within this URL was the user's access token and client ID. Also of interest in the filtered Dominator Tree was the URL for accessing the user's notebooks and the GUID of the user's parent notebook.

By executing an OQL query to find all URLs within the memory, we were able to locate URLs to notebooks and notebook sections that had been stored in memory and, thus, likely to have been most recently accessed by the user. OneNote does not seem to store user information, such as the user's email address or username, directly within the memory.

As mentioned in Section 2.6.4, the refresh token is stored more securely than that of Microsoft OneDrive's, but it is merely encrypted with data residing upon the device. The authentication scheme appears identical to that of OneDrive's in that the refresh token must first be used to generate an access token from the Microsoft Live servers before using this received token to authenticate all other file access.

### 2.8.7 Universal Password Manager Analysis

A HPROF memory heap was obtained from the UPM app. The file was analyzed with Eclipse Memory Analyzer and filtered to show only objects that were within UPM's package name ("com.u17od.upm").

Firstly, we were able to locate instances of UPM's crypto classes. Within these classes we located the 8-character salt used by UPM to encrypt and decrypt the database containing the user's entries. Furthermore, we were able to locate the cipher used by UPM ("PBEWithSHA256And256BitAES-CBC-BC"). We proceeded to search the entirety of the memory and located an instance of the "javax.crypto.spec.PBEKeySpec" class. This Java class contained the user's password in plaintext. The use of OQL queries further strengthened the results we found via filtering of the Dominator Tree.

As previously discussed in Section 2.7.4, UPM does not communicate directly with a server. As such, the only methods to obtain the data within the UPM database are brute-force techniques based on the findings above and in Section 2.7, and obtaining the user's password via other means.

## 2.9 OUR RESEARCH ENVIRONMENT

In our experiments, we used several physical Android devices and an Android virtual machine. We used two popular Android phones: the Nexus 4 (E960) and Samsung Galaxy S3 (i9300). During our experiments, the Nexus 4 was running under both Android 4.2.2 and Android 4.4. The Galaxy S3 was running Android 4.0.4 and the virtual machine ran a version of Android 4.4.

The following describes the device specific procedures we performed with the above devices.

### 2.9.1 Nexus 4

The Nexus 4 was used as our primary device for validating our proposed evidence collection and analysis methodology [1]. The following describes the process we undertook in order to prepare the device for the evidence collection, and examination and analysis.

When we first obtained the device, the Nexus 4 was running Android 4.2.2. This version of Android was released in early 2013, with many high-end phones being released towards the end of 2013 still running this version of Android. Using Android 4.2.2 on the Nexus 4 device, we proceeded to unlock the bootloader (see [1] for further details), causing Android to display a warning indicating that all data on the device would be wiped upon unlocking of the bootloader. We proceeded with unlocking the bootloader and continued to load, within the device's RAM, a custom recovery image designed specifically for the Nexus 4.

The recovery mode successfully booted, allowing us to access the device (with full read and write privileges) via the ADB application over a USB cable. Navigating into the device's "userdata" partition showed that no files had yet been wiped; all user files were intact. Rebooting the device at this stage causes the Android device to wipe this partition. Using the ADB application, we successfully performed a "bit-for-bit" copy of all the device's partitions. We proceeded to modify the "services.jar" file to allow for our specially signed app to access user account information.

The Nexus 4 stores all "classes.dex" files for its system package's JAR files within the JAR file. This made it much easier for us to perform the required code injection on this file (as opposed to the Galaxy S3, see Section 2.9.3). We retrieved the "services.jar" file from the "/system" partition while in the custom recovery mode, injected our code to allow APKs containing our signature to bypass signature checks, and then replaced it on the device. Deleting the "services.odex" file on the device then prompts the device to rebuild all system packages upon next startup. No further change was necessary to allow our signature to have system level privileges on the device. Following this, we analyzed the apps (via their forensic images). In order to read the accounts data, we copied our APK file into the "/system/app" location and rebooted the device. As our APK was signed such that it could imitate any package on the device, we were able to retrieve all accounts data from the Nexus 4's secure credentials storage.

Following our evidence collection and analysis methodology, we noted that upon encryption of the device, the "userdata" partition is encrypted but the "system" partition is not. This means that our above methodology could be utilized to inject code into any system package even on an encrypted device. We also note that Android versions after 4.2.2 erase the "userdata" partition upon unlocking of the bootloader. This means a different method would be required in order to unlock the bootloader to run the custom live OS.

### 2.9.2 Android VM

Our Android (user debug build) virtual machine was used primarily for the app analysis stage of our evidence collection and analysis methodology. By using a user debug build of the Android OS, we were able to obtain memory heaps (HPROF files) of all apps running regardless of whether they explicitly allow or disallow this action. To a forensic practitioner, this is an immensely useful tool to have as by default, Android apps do not allow the OS to obtain memory heaps of their data unless the app's "AndroidManifest.xml" file contains the "android:debuggable='true'" directive.

By utilizing the Android virtual machine, we were able to conduct in-depth analysis of the apps which were examined in Sections 2.1 to 2.7. By employing the aforementioned HPROF memory heaps, we obtained the methods in which the data obtained in the initial analysis of the apps could be used in order to authenticate as the user (see Section 2.8). Without the information from each app's memory, it would be unlikely a practitioner could succeed in authenticating as these apps.

### 2.9.3 Samsung Galaxy S3

In our experiments, the Samsung Galaxy S3 was utilized as a secondary device used mainly for device specific sections of our research. Most notably, when we extracted the "services.jar" file from its "system/framework" directory, we found that system packages on the Galaxy S3 do not contain a "classes.dex" file. This means we were required to go through an alternate route of repackaging each dependent JAR and APK file in the "system/framework" and "system/app" directories respectively such that the respective "odex" files were deodexed and placed as "classes.dex" files within the APK or JAR. After the JAR and APK files were replaced, we deleted their matching "odex" files in order to trigger the Android OS to rebuild the "odex" files from each app or framework's package.

The reason this process is required is due to Samsung electing to store only each system framework or app's source code within the optimized "odex" form of each framework or app. By default, the Android OS keeps all "classes.dex" files within the main package. The method Samsung employed reduces the amount of space that the "system" partition takes up significantly. In our experiments, in order to replace the "services.jar" file with our own modified version, we were required to replace a total of 175 items, with 56 of those belonging in the "system/framework" directory and the remaining 119 in the "system/app" directory. We accomplished this using a script that monitored "logcat" output for occurrences of "StaleDexCacheError". This resulted in a successful implementation of our methodology, giving us access to the entire AccountManager API.

## 3 CONCLUSION

In this chapter, we implemented the methodology proposed in [1] using six popular cloud apps and one password sync app as a case study to determine the types of forensic artifacts that can be collected from Android devices.

We found that all four cloud storage apps saved cached or offline files on the device's external storage. While some of the apps deleted the files after a period of time (once they were no longer being used) when the files were allocated, the majority of files were not protected. One exception to this was Box, which encrypted the files, requiring an encryption key and salt kept in the app's private storage directory to decrypt. Cloud storage apps also commonly kept thumbnails and other preview files on the device external storage. On internal storage, these apps generally stored file metadata (both for files cached on the device and files stored on the server) in an SQLite database. The usefulness of this metadata for forensic purposes varied between apps. A large amount of metadata was found, which could be of use in forensic investigations (including timestamp and file hash information). Cloud apps generally use tokens to authenticate the connection with their remote cloud servers. We found that these tokens were stored in a varied number of formats (although most of the authentication schemes were based on OAuth 1.0 or 2.0), which often required further analysis to decode or decrypt. App configuration files were also shown to contain some data of forensic interest, often this was related to authentication or encryption; however, user metadata (e.g. account names, emails) was also commonly found.

The two note taking apps shared many similarities with the cloud storage apps, leading us to the conclusion that most cloud based file-focused apps store similar types of data. However, the format and structure of the data is generally different. Microsoft's OneNote, for example, did not store any files on external storage, instead storing our test note files only on private internal storage. Generally the note taking apps often treated (and stored) notes as individual files or in combined files (such as Notebooks). The formats of these files were not particularly complex with Evernote using an XML based format, which made examination straightforward and OneNote using the standard OneNote file format suitable for opening in the PC version of the app.

We found that using the information obtained from both the initial and further analysis sections, a practitioner could access the cloud service's servers as the user (and access their files) on the device for five of the six apps we tested that communicated and authenticated directly with cloud services. These apps were Dropbox, Box, OneDrive, ownCloud and OneNote.

Future work includes advanced analysis of device bootloaders to determine techniques of booting live operating systems, where this is not supported by manufacturers by default, to facilitate forensic collection.

## REFERENCES

[1]   B. Martini, Q. Do, and K.-K. R. Choo, "Conceptual Evidence Collection and Analysis Methodology for Android Devices," In Ko R and Choo K-K R, editors, Cloud Security Ecosystem, Syngress, an Imprint of Elsevier, 2015.
[2]   K. Goundan. "Using OAuth in "PLAINTEXT" mode " 25th July 2014; https://www.dropbox.com/developers/blog/20/using-oauth-in-plaintext-mode